# Modular approach to microswimming

Ran Niu*, Thomas Palberg

Institut für Physik, Johannes Gutenberg-Universtät Mainz, Staudingerweg 7, 55128 Mainz, Germany

## Abstract

The field of active matter in general and microswimming in particular has experienced a rapid and ongoing expansion over the last decade. A particular interesting aspect is provided by artificial autonomous microswimmers constructed from individual active and inactive functional components into self-propelling complexes. Such modular microswimmers may exhibit directed motion not seen for each individual component. In this review, we focus on the establishment and recent developments in the modular approach to microswimming. We introduce the bound and dynamic prototypes, show mechanisms and types of modular swimming and discuss approaches to control the direction and speed of modular microswimmers. We conclude by highlighting some challenges faced by researchers as well as promising directions for future research in the realm of modular swimming.

## 1. Introduction

Scientists have been gaining enormous inspirations from biology in designing smart devices and nanomachines. The swimming microorganisms (e.g., bacteria, sperms and cells) have triggered the design of artificial micro-swimmers (~1-100 μm in size), which are capable of self-propulsion at low Reynold's number by converting energy from one form (e.g., chemical) to the other (e.g., mechanical).[1-6] The propulsion of artificial swimmers can be phoretic, acoustic or bubble thrust. Taking the well explored phoretic swimmers as an example, phoretic propulsion relies on local gradient of some kind such as electric potential (electrophoresis), concentration of solute (diffusiophoresis) and temperature (thermophoresis).[7-11] To break time reciprocity, swimmers of asymmetric shapes (e.g., ellipsoid, rod, star, fish-like) or surface properties (e.g. patchy)[12-16] have been manufactured thanks to the advances in micro- and nanofabrication techniques.[17-19] Alternatively, artificial swimmers explore radically different environments, e.g., periodic light field and external gradients of some kind for emergent behaviors,[20-23] just like the chemotaxis of bacteria and sperms as a way to avoid toxins or to browse for nutrients. The assemblies of artificial swimmers also feature similar collective behaviors as what happen in natural world, for instance, flocking migration of birds, swarming and swirl schooling of fish.[24-29] Furthermore, they have been developed for a wide variety of applications, including motion-based active sensors, cargo transport, targeted drug delivery, as well as environmental monitoring and remediation.[30-35]

However, most of the above introduced swimmers are based on single active unit and combine all relevant functions in the same single entity. While biology obtain complex structure from assembly of different components, for example, the pairing of DNA, the assembly of protein chains into quaternary structures and the formation of cells from tissues. Can we also construct artificial swimmers from different components? The answer is yes. Whitesides and coauthors first demonstrated the assembly of millimeter scale hemicylinderical plates of the same or opposite chirality into homo- or heterochiral dimers, which perform directed motion not seen by individual plates.[36] Since then, many artificial modular swimmers, which assemble different active and inactive components into self-propelling complexes, have been designed and realized. Two categories of modular prototypes, i.e., bound and dynamic swimmers have been made. A large toolbox has been explored in assembly of modular prototypes, including passive spherical particles (polystyrene, silica, etc.), arbitrary shaped (e.g., dumbbell, rod, plate-like) passive particles and functional particles (e.g., superparamagnetic/ferromagnetic particles and patchy particles). The distinctive and indeed defining feature of modular prototypes is the multiple interacting and cooperative components, which leads to flexibility in combination, assembly, functionality and optimization for better performance. Therefore, modular approach greatly relieves the demands for very sophisticated fabrication of swimmers or construction of environments to obtain propulsion, steering or targeted cargo release. The significant research activity in constructing modular swimmers during the past decade motivates this review.

In this article, we first review and classify the modular prototypes introduced in the literature according to the flexibility of structure and the mechanism of assembly, i.e., whether it relies on external fields or self-generated gradient/field. Then we summarize the mechanisms for self-propulsion and the types of modular swimming. The multicomponent character allows modular swimmers to combine different functional components for steering. In section 4, we introduce the various approaches to steering modular swimmers for direction and speed control. We discuss the experimental and theoretical challenges confronted by modular approach in section 5, and close by discussing possible directions for future investigations.

## 2. Modular assembly of modular swimmers

In the literature, mainly two kinds of modular swimmers have been reported, i.e., swimmers with bound and dynamic structures. For bound structures, all the components are linked by chemical bonds or electro-static interactions, therefore the components lose their rotational degree of freedom. On the contrary, in dynamic structures, each component has some rotational degrees of freedom. Thus, dynamic structures are more flexible as the composition and organization of components can be changed online. Moreover, dynamic structures can rearrange, disassemble and re-assemble in response to physicochemical stimuli. Furthermore, the assembly of dynamic structures can be induced by external fields or self-generated gradient/field.

### 2.1 Modular swimmers with bound structures

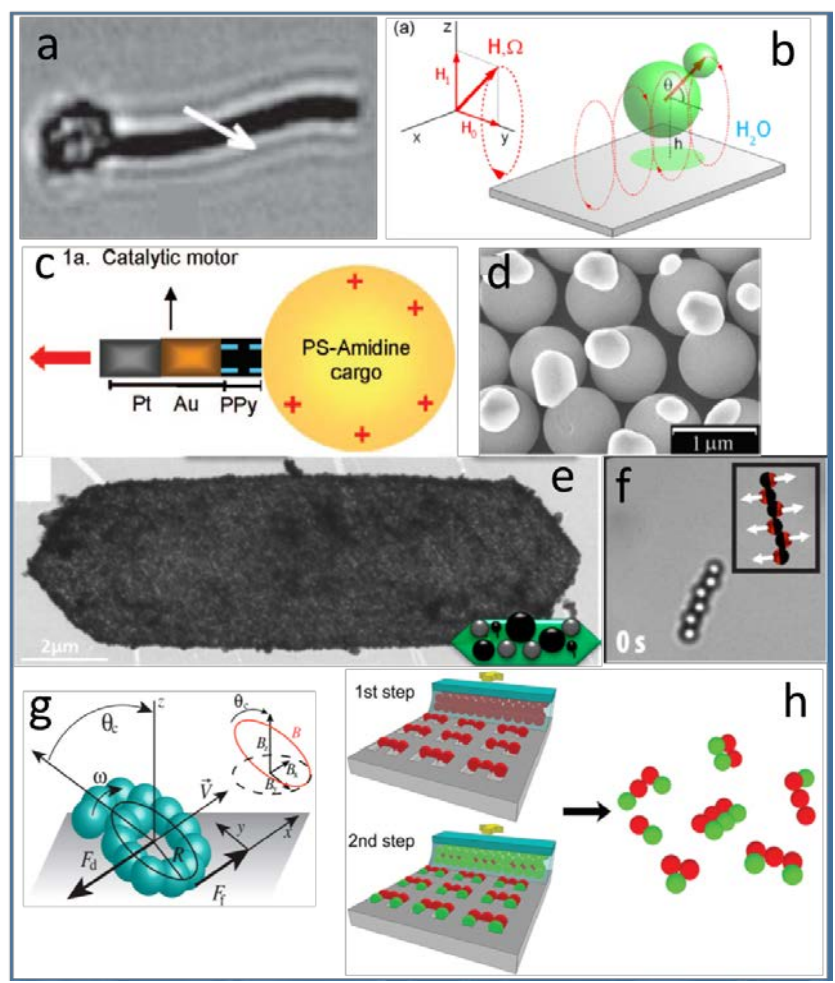

**Fig. 1** Examples of modular swimmers with bound structures in chronological order. (a) A DNA connected filament of magnetic particles linked to a red blood cell by biotin-streptavidin interaction.[37] (b) DNA linked anisotropic doublet composed of asymmetric paramagnetic particles.[38] (c) Colloidal cargo linked to a tri-segmented nanorod by electrostatic interaction.[39] (d) Catalytic dimer of platinum particle on top of a silica sphere fabricated by deposition and thermal annealing.[40] (e) Active nanoparticles ($Fe_3O_4$, Pt or gold) chemically bound to a polymer single crystal of nanometer thickness.[41] (f) Colloidal chain made of Janus particles with 'zig-zag' shaped arrangement.[42] (g) Magnetic lasso composed of superparamagnetic particles flexibly linked by entanglement of adsorbed polymer between adjacent particles with the aid of modest heating.[43] (h) Colloidal molecules obtained by sequential capillary-assisted assembly.[44] (Adapted (a) from Ref. 37, Copyright © 2005, Springer Nature. Reprinted (b) with permission from P. Tierno, *et al.*, *Phys. Rev. Lett.*, 2008, **101**, 218304. Copyright 2008 by the American Physical Society. Reprinted (c) with permission from *Nano Lett.*, 2008, **8**, 1271. Copyright 2008 by American Chemical Society. Reprinted (d) with permission from Ref. 40. Copyright © 2010 Wiley. (e) Reprinted with permission from *ACS Nano*, 2013, **7**, 5192.

The first bound structure was fabricated by Dreyfus et al. Mimicking flagella of bacteria, they made artificial flagellum of a DNA bound flexible magnetic chain, which is fixed to a red blood cell to achieve non-reciprocal motion when actuated by external magnetic fields (Fig. 1a).[37] Tierno et al. connected differently sized paramagnetic particles into anisotropic doublets for controlled propulsion when floating on a flat surface and subjected to a magnetic field processing around an axis parallel to the surface (Fig. 1b).[38,45] Their modular design includes the doublet and the charged substrate for asymmetric dissipation during cyclic rotation. Following this work, Cheang et al. constructed magnetic triplets of bended configuration self-propelling in spatially uniform, time dependent magnetic fields.[46] Sundararajan et al. tethered a prototype cargo to one end of a bi- or tri-segmented rod by biotin-streptavidin or electrostatic interactions (Fig. 1c).[39] Later on, the authors designed a dissolvable binding via photochemical stimuli to release the cargo on demand.[47] Valadares and coworkers proposed a simple way to fabricate catalytic dimers by depositing a platinum layer on top of a silica sphere followed by thermal annealing induced de-wetting and formation of an attached platinum sphere over the silica sphere (Fig. 1d).[40] This method can fabricate a large amount of dimers in one batch. Dong et al. chemically bound thousands of functional nanoparticles of different species ($Fe_2O_3$, platinum and Au) on a polymer single crystal of nanometer thickness (Fig. 1e).[41] This method enables multiple propulsion mechanisms and enriches the functionalities available for one swimmer. Besides Dreyfus and coworkers, there are a few groups interested in connecting particles into colloidal chains of different rigidities.[42,43,48] We specially mention the colloidal chains fabricated by Vutukuri and coworkers of 'zig-zag' configuration (Fig. 1f). The colloidal chains were obtained by assembly of Janus particles inside an electric field followed by heat-induced crosslinking. The 'zig-zag' configuration induces a constant torque acting on the chain, therefore the rotational motion of the chain. Moreover, the rigidity of the chain is tunable by the chemical composition of the polymer spheres used for making Janus particles.[42] Another nice example is the flexible superparamagnetic chains made by Tang and coworkers in a planar rotating magnetic field (Fig. 1g).[43] The chain is so flexible that it is deformable to trap cargo inside the chain and release the cargo on demand.

Ni et al. sequentially assembled microparticles of different kinds into hybrid colloidal molecules of prescribed shapes in fabricated templates followed by linking via sintering (Fig. 1h).[44] The arrangement of particles inside the molecule is controlled by the sequence of feeding. Therefore, the advantage of this method is a high programmability in the choice of material, composition and shape.

## 2.2 Modular swimmers of dynamic structure

### 2.2.1 Modular assembly driven by external fields

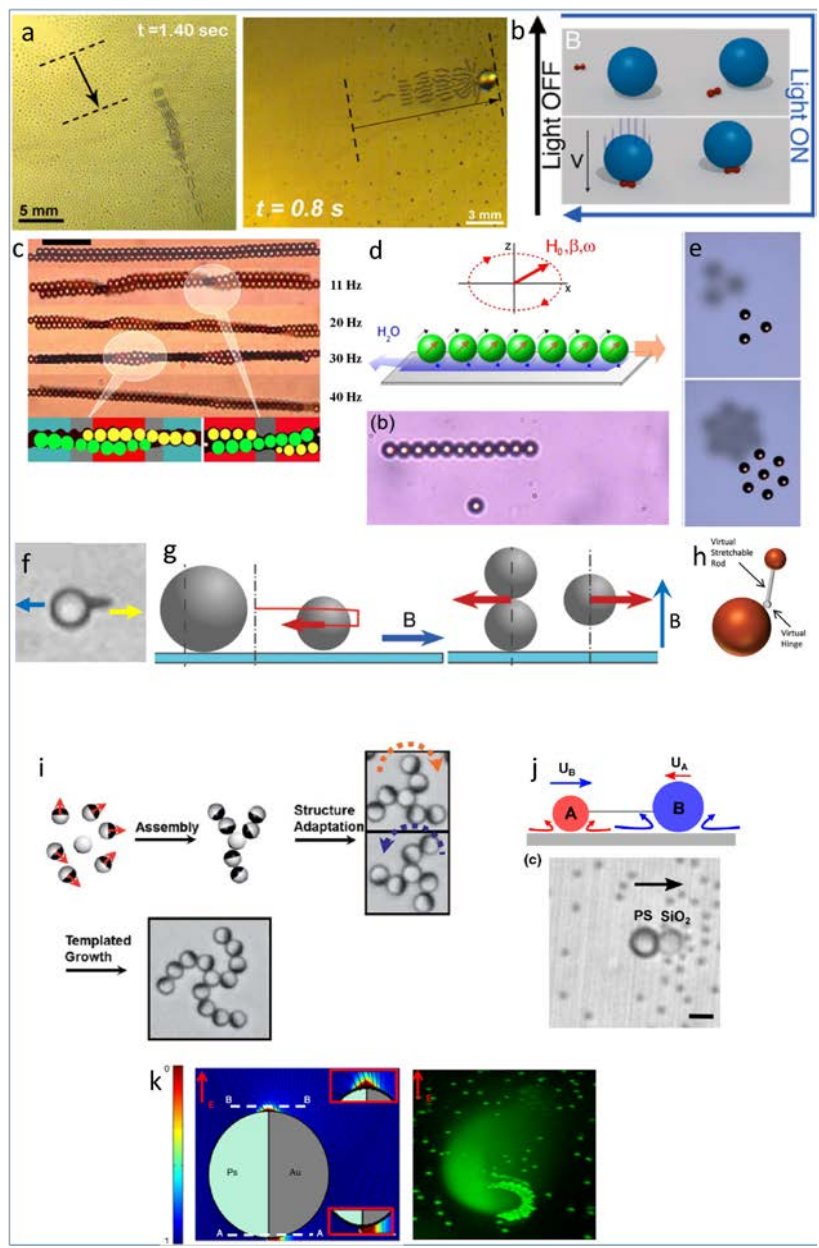

**Fig. 2** Modular swimmers of dynamic structures self-assembled under external magnetic (a-h) or electric fields (i-k). (a) Self-propelling snake at high frequency magnetic field (left) and snake-particle hybrid without increasing the field frequency (right).[49] (b) Peanut shaped hematite is guided by weak magnetic field to the vicinity of the colloid, where they couple by light activated phoretic force from a chemical gradient.[50] (c) Helical ribbon formed by ensemble of paramagnetic beads in external magnetic field.[51] (d) Propelling worm of magnetic rotors formed on a glass plate and subjected to an elliptically polarized rotating magnetic field.[52] (e) "Crystals" of ferromagnetic beads formed by magnetic Cheerio effect at air-liquid interface.[53] (f) Sphere-nanorod prototype assembled in an external rotating magnetic field.[54] (g) Thrower and rower formed of asymmetric paramagnetic beads actuated by external periodic magnetic fields.[55] (h) Two-body swimmer formed by paramagnetic beads under eccentric magnetic field.[56] (i) Rotating pinwheel formed by Janus particles centered at a homogeneous colloid under an AC electric field.[57] (j) Asymmetric colloidal dimer coupled by unbalanced electrohydrodynamic flow near a conducting surface in a vertically applied electric field.[58] (k) Janus particle selectively attracts or repels colloids by dielectrophoresis in a vertical electric field.[59] (Reprinted (a) with permission from A. Snezhko, *et al.*, *Phys. Rev. Lett.*, 2009, **102**, 118103. Copyright 2009 by the American Physical Society. Reprinted (b) with permission from *J. Am. Chem. Soc.*, 2013, **135**, 15978. Copyright 2013 American Chemical Society. Reprinted (c) with permission from N. Casic, *et al.*, *Phys. Rev. Lett.*, 2013, **110**, 168302. Copyright 2013 by the American Physical Society. Reprinted (d) with permission from F. Martinez-Pedrero, *et al.*, *Phys. Rev. Lett.*, 2015, **115**,138301. Copyright 2015 by the American Physical Society. Reproduced (e) from Ref. 53, http://creativecommons.org/licenses/by/4.0/. Reprinted (j) with permission from F. Ma, *et al.*, *Phys. Rev. Lett.*, 2015, **115**, 208302. Copyright 2015 by the American Physical Society.)

External magnetic/electric fields have shown their advantages in the construction of dynamic modular swimmers (Fig. 2) with on/off functionality and better controllability under the expenses of designing a well-controlled field and consumption of energy. Generally, magnetic field can induce magnetic and hydrodynamic interactions between components or navigate anisotropic magnetic particles for assembly. Snezhko et al. reported that the coupling between surface deformation of the fluid at air-liquid interface and the collective response of magnetic particles to an external alternating magnetic field induces a stable snake of quadruple flow.[60,61] The snake is able to self-propel at high frequency field (above a transition region, left in Fig. 2a). Alternatively, without increasing the frequency of the magnetic field, assembly of a glass bead at one end of the immobile magnetic snake suppresses the votex flow at the tail giving rise to uncompensated flow that propels the swimmer (right in Fig. 2a).[49] Palacci et al. utilized a weak uniform magnetic field to guide peanut shaped magnetic docker to the vicinity of a cargo particle, where they couple by phoretic force resulting from a chemical gradient generated by a light activated catalytic reaction (Fig. 2b).[50] The docker can couple different species of cargo particles. Moreover, it is reusable and can thus accumulate cargo at a targeted place. Casic et al. assembled helical ribbons following the steps of chain formation, ribbon formation and fragmentation, segment healing and coalesce by combination of dipole interactions and gravity force (Fig. 2c). The structure is interesting; however, the control using different magnetic fields in each step is complicated. Martinez-Pedrero and coauthors dynamically assembled propelling worms of magnetic rotors close to a confining glass plate and subjected to elliptically polarized rotating magnetic field due to the cooperative flow generated by the spinning particle acting as a hydrodynamic "conveyor belt" (Fig. 2d).[52] At higher particle concentration, the authors showed the formation of dynamic carpets by the same effect.[62]

Lumay et al. assembled small "crystals" of ferromagnetic beads floating at air-fluid interface and subjected to a vertical magnetic field induced repulsive dipole-dipole interaction, the so-called magneto Cheerio effect (Fig. 2e).[53] These "crystals" are actuated by a horizontal and oscillating magnetic field into pulsating mode. Above a field strength, a translational component can be triggered by associating hydrodynamic coupling and a non-reciprocal deformation of the swimming body.[63] Inspired by this design, García-Torres et al. and Vilfan et al. designed rod-sphere and asymmetric modular prototypes (Figs. 2f and 2g). The rod-sphere prototype composes of a ferromagnetic nanorod and paramagnetic sphere in an in-plane square wave modulated magnetic field, by which the pair dipole interaction of constituents give rise to periodic attraction-repulsion sequences (Fig. 2f). And the symmetry breaking induced by the coordinated interaction of the constituents leads to directed motion. Depending on the strength of the field, the nanorod can both drag and push the paramagnetic particle. As well as the swimmers shown in Figs. 2b, 2d and 2e, this prototype shows well-controlled directionality. Additionally, it can realize multicargo transport. Vilfan et al. assembled asymmetric paramagnetic beads, i.e., differently sized beads (thrower) or one bead and a dumbbell (rower) in periodic horizontal-vertical oriented magnetic field (Fig. 2g). The field induces periodic attraction and repulsion between components, which in combination with gravity, the proximity to a non-slip boundary or re-rotation of anisotropic components breaks the time reciprocity of motion. Although similar in assembly and propulsion mechanisms to the "crystals" and rod-sphere prototypes, the thrower and rower are omnidirectional, meaning that the direction of motion is not determined by the magnetic field but the orientation of the swimmer. Therefore, this prototype is more autonomous. Very recently, Du et al. showed two-body and multibody swimmers formed by paramagnetic particles in eccentric magnetic field with strong anisotropy to break time reciprocity.[56]

In principle, external electric field generates dipole-dipole interaction or induced-charge electro-osmosis for assembly of modular swimmers. Zhang et al. reported the self-assembly of Janus particles into rotating pinwheels when mixed with trace amount of homogeneous colloids in the presence of an AC electric field, resulting from the attractive dipole-dipole interaction. The pinwheel breaks the symmetry in rotational direction producing spiral and chiral shapes (Fig. 2i).[57] Each arm is flexible and stable as the head-to-tail arrangement of Janus particles is fulfilled. Ma et al. assembled asymmetric particles of different materials, sizes or surface charges into self-propelling dimers via the tangential electrohydrodynamic (EHD) flow generated on a charged surface in a vertical electric field (Fig. 2j).[58] In principle, any kind of asymmetric particles can be used, thus this approach greatly enriches the toolbox for dynamic modular assembly. Recently, Boymelgreen and coworkers utilize metallodielectric Janus particle with reverse moving direction (i.e., metal side in the front) to attract or repel colloidal particles of unique frequency response by self-dielectrophoresis in a vertical electric field (Fig. 2k).[59] Consequently, this method has dynamic control of particle selection.

*2.2.2 Modular self-assembly driven by self-generated gradient/field*

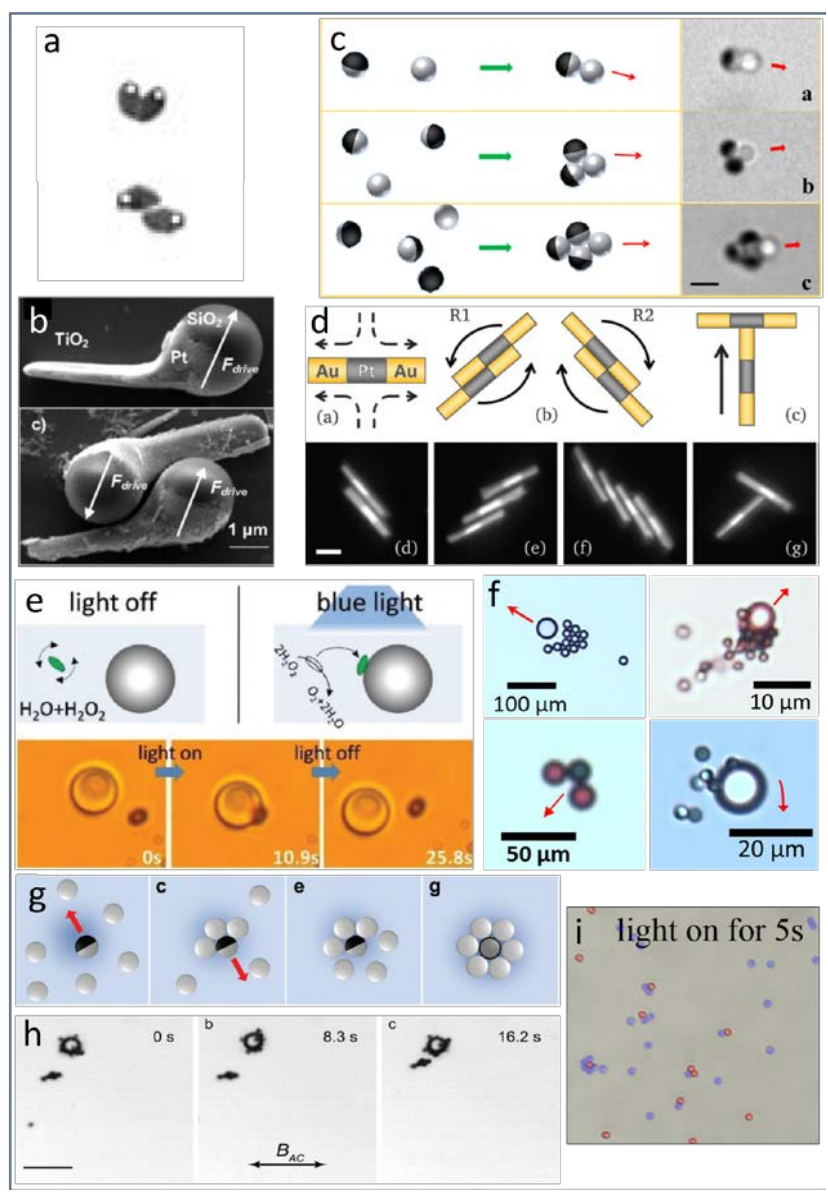

**Fig. 3** Different strategies for self-assembly of dynamic modular swimmers driven by self-generated gradient/field in chronological order. (a) Heterochiral and homochiral dimers formed by hemicylindrical plates of the opposite and same chirality, respectively.[36] (b) Two tadpoles self-assemble into a cluster bound by van der Waals interaction.[64] (c) Assemblies of Janus motors and non-catalytic hydrophobic colloid coupled by hydrophobic interaction.[65] (d-g) Modular swimmers self-assembled by hydrodynamic force arising from catalytic reactions (d, e and g)[66-68] and ion exchange induced concentration gradient (f).[69] (h) Self-assembly of ferromagnetic floaters by magnetic attraction at air-fluid interface into a "snake".[70] (i) Active colloidal molecules assembled from heat adsorbing and non-adsorbing particles by self-diffusiophoresis.[71] (Reprinted (a) from Ref. 36. Copyright © 2002 WILEY. Reprinted (b) from Ref. 64. Copyright © 2010 Wiley. Reprinted (c) with permission from *J. Am. Chem. Soc.*, 2013,

Alternatively, modular swimmers can also self-assemble from self-generated gradient/field. Several successful prototypes have been realized (Fig. 3). Whitesides and coauthors first showed the assembly of millimeter scale hemicylindrical plates of the opposite and same chirality into heterochiral and homochiral dimers driven by surface tension in 2002 (Fig. 3a).[36] The heterochiral dimers move unidirectionally in contrast to the rotational motion of individual plates. Gibbs et al. reported the assembly of two tadpoles into a rotating cluster assisted by random Brownian motion or active motion resulting from the catalytic reaction of single tadpole (Fig. 3b).[64] The authors refer to van der Waals interaction as the main mechanism for stabilizing the two tadpoles. Gao et al. demonstrated the assembly of multiple catalytic Janus motors and nearby homogeneous hydrophobic colloids into self-propelling swimmers dynamically coupled by hydrophobic interaction (Fig. 3c).[65] The self-propulsion of Janus motors increases the chance of assembly, however, it is not determinative to the directed motion of the cluster. Without the hydrophobic colloid, clusters of Janus motors only have rotational motion, proving the modular feature of the Janus-sphere prototype.

Another prominent and important strategy for self-assembly is long-ranged hydrodynamic interaction (Figs. 3d-g), which decays with $1/r$ , $1/r^2$ or $1/r^3$ from the source depending on whether it is generated by a driven or force-free object and also on the cell geometry.[72-74] A few groups independently demonstrated the self-assembly of bi- or tri-segmented metallic nanorods into various configurations, e.g., staggered doublets or triplets and T-swimmers in a chemically active media driven simultaneously by electrostatic and hydrodynamic forces (Fig. 3d).[66,75,76] Following the work of Palacci et al,[50] Martinez-Pedrero et al. synthesized ellipsoid shaped hematite dockers, which at high fuel concentration attract passive particles within several micrometer range activated by blue light (Fig. 3e). After docking by osmotic force, the composite self-propels by diffusiophoresis at low speed. The authors further applied a rotating magnetic field as an independent mean to propel the whole composite. This prototype can combine multiple dockers and multiple passive particles and release the components by switching off the illumination.

We have been utilizing ion-exchange induced pH gradient by an ion exchange resin (IEX) to generate millimeter ranged converging electro-osmotic (eo) flow on a negatively charged substrate which decays as $1/r$ in confined geometry and $1/r^2$ when it is not.[77] The eo-flow drifts colloidal particles to the vicinity of the central IEX and couple with it dynamically forming a modular swimmer. Therefore, our modular construction includes the IEX, many charged colloidal particles and a charged substrate. A distinctive feature of the IEX-based modular swimmers is that they self-propel in low ionic and electro-neutral suspension. Thus, the eo-flow is rather long-ranged and the assembled colloids keep some distance between each other allowing for structural reorganization. Moreover, the eo-flow can transport and couple a wide variety of charged particles e.g., passive colloidal particles, superparamagnetic particles, anionic ion exchange resin particles and even Janus particles (Fig. 3f). Furthermore, the swimmers can also assemble multiple IEX with many passive particles.

Singh et al. demonstrated the assembly of silica-titanium dioxide Janus sphere and nearby passive particles by diffusiophoretic attraction initiated by UV light. This prototype can transport both negatively and positively charged colloids at near neutral pH conditions. Once they couple by phoretic force, the newly formed swimmer drops its velocity. Moreover, it reverses the moving direction as a counter flow from the assembled particle opposes the osmotic flow over the silica surface of the Janus particle (Fig. 3g).[68] More general aspects on hydrodynamic force induced manipulation and transport of colloids can be found in a recent review.[78]

Kokot and coworkers showed the self-assembly of ferromagnetic floaters into functional structures such as rings and rods at air-liquid interface owing to the magnetic interactions (Fig. 3h).[70] It should be noted that the self-assembled structures do not self-propel, they need external oscillating magnetic field for long-range magnetic and hydrodynamic interactions on the fly. Recently, Schmidt et al. presented the self-assembly of heat adsorbing and non-adsorbing particles in near critical phase separation suspension triggered by the illumination of light (Fig. 3i). The mechanism for assembly is concentration gradient induced self-diffusiophoretic force within some distance range.[71]

From design point of view, modular swimmers shown in Figs. 3f, 3h and 3i are completely relieved from surface coating or asymmetric shape for propulsion. In particular, the prototypes shown in Figs. 3f and 3i, although assembled by different physical mechanisms, act as good model systems verifying the feasibility of the simple theoretical model proposed by Soto and coworkers[79,80] that particles of different surface activities and mobilities can self-assemble into self-propelling swimmers. From functionality perspective, the swimmers shown in Figs. 3e, 3g and 3i have on/off switch, and the swimmers in Figs. 3e and 3f (upright) can be steered by external magnetic field.

## 3. Mechanisms and types of modular swimming

### 3.1 Mechanism of self-propulsion

In general, modular swimming can be achieved by two main propulsion mechanisms, i.e., powered by local conversion of energy or driven by external fields. Self-powered modular microswimmers include a gradient/field generating component for phoretic propulsion. The gradient/field generating component can be catalytically active, e.g., platinum or titanium dioxide deposited or hematite embedding particles (Figs. 1c-f, 2a, 3a-e, 3f and 3h). Therefore, as other catalytic swimmers, these modular swimmers also depend on neutral and ionic-diffusiophoretic as well as electrophoretic propulsion.[81-83] Other non-catalytic modular phoretic swimmers rely on ion exchange induced pH gradient (Fig. 3f) and critical phase separation induced concentration gradient (Fig. 3i) for self-propulsion.[69,71,84] Our group has been working on ion exchange induced pH gradient and flow for self-assembly and modular swimming.[69,77,85-88] Our current understanding of the self-propulsion mechanism is the overall flow symmetry breaking from two main sources. Generally, the first assembled colloid actively breaks the radially symmetric flow generated by the IEX and sets the swimmer into motion. Once the swimmer moves, the pH gradient generated by the IEX becomes asymmetric, leading to asymmetric eo-flow.[69] However, the contributions of diffusiophoresis, electrophoresis and chemophoresis to the propulsion are not clear for the time being. A few groups have worked on the behaviors of Janus swimmers with carbon coated hemisphere in near critical binary mixture activated by light.[27,89-91] Schmidt and coworkers took a step further to separate the heat adsorbing and non-adsorbing hemispheres into two species of particles, which thus made the swimmer modular. The attachment of a heat non-adsorbing particle to a heat-adsorbing particle generates a temperature gradient, which gives rise to a chemical gradient for self-diffusiophoresis.[71]

The second category of modular swimmers are actuated by uniform external fields, i.e., they are driven out of equilibrium by external fields. Magnetic field actuated modular swimmers are shown in Figs. 1a, 1b, 1g, 2a-h and 3h. In general, magnetic fields either mechanically generate asymmetric hydrodynamic flow or induce non-reciprocal deformation of the swimmer body for propulsion. When placed in a uniform magnetic field, magnetically responsive particles subjected to magnetic torque align their easy axis with the direction of field. Therefore, a rotating or oscillating magnetic field generates particle motion below a critical frequency. When the structure of the modular swimmer is asymmetric or the magnetic functional component is anisotropic, the magnetic field leads to directed motion (e.g., Figs. 1b, 2a-d, 3e, 3h). Alternatively, oscillating magnetic field induces periodic attraction and repulsion between components for swimmers shown in Figs. 2e-h, which together with viscous effects on a non-slip surface generates non-reciprocal motion. Modular swimmers prompted by external electric fields are shown in Figs. 1h, 2i-k, which rely on the generation of unbalanced flow from induced charges around the complexes for propulsion. Breaking the symmetry in polarity of complex or shape induces self-propulsion. Interested readers are referred to a recent review[92] summarizing progresses on engineering microbots and microdevices powered by magnetic and electric fields.

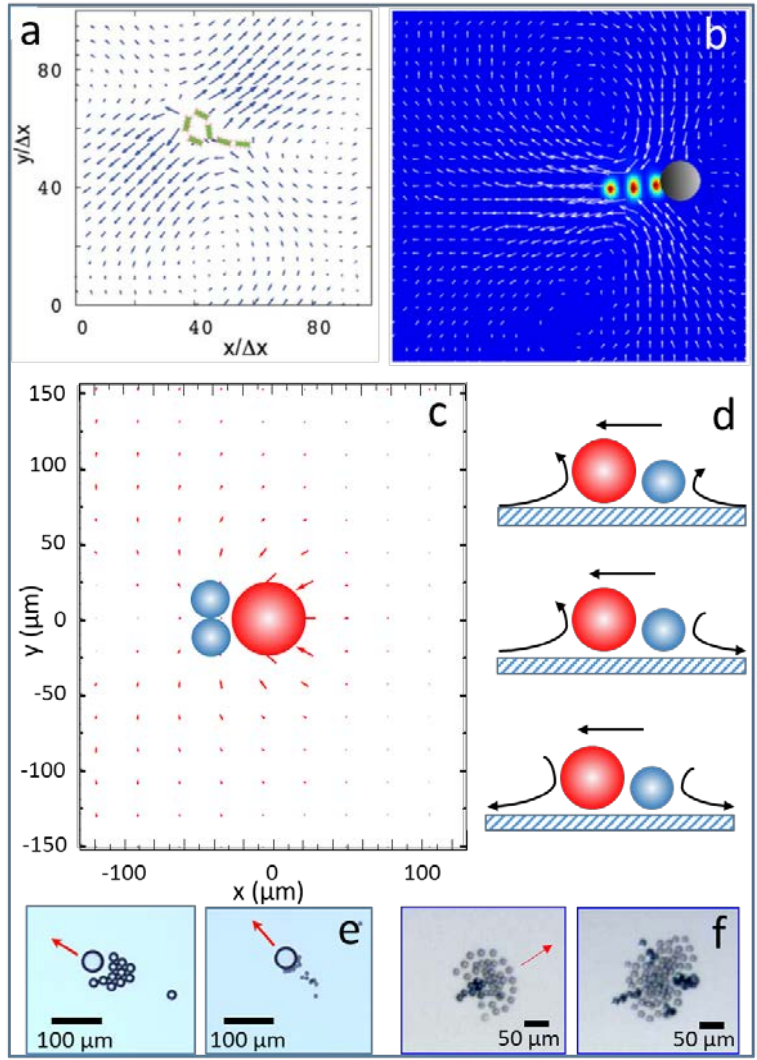

**Fig. 4** Typical flow patterns around modular swimmers (a-d). (a) Hydrodynamic flow induced by a self-propelling magnetic "snake" in an oscillating magnetic field.[70] (b) Velocity field of a particle-snake hybrid in external magnetic fields.[49] (c) Asymmetric flow generated by modular swimmer composed of single IEX (red) and two passive colloids (blue) on the bottom substrate (top view in the co-moving frame). (d) Schematics of typical 3D flow pattern generated by IEX-based modular swimmers and asymmetric dimers in EHD flow (side view). (e) Modular swimmers formed by single IEX and several colloidal particles of large (left) and small sizes (right). (f) Modular swimming of multiple IEX and multiple colloidal particles distributing asymmetrically (left) and eo-pumping of symmetrically distributed colloidal particles around the cluster of IEX (right). (Reproduced (a) from Ref. 70, http://creativecommons.org/licenses/by/4.0/. Reprinted (b) with permission from A. Snezhko, *et al.*, *Phys. Rev. Lett.*, 2009, **102**, 118103. Copyright 2009 by the American Physical Society.)

Most modular swimmers are capable of perturbing the surrounding fluid in a way that generates a net displacement of their own body. Moreover, the whole complex generates complicated flow pattern travelling along with it. Fig. 4 shows typical flow patterns under modular swimming. When actuated by an alternating magnetic field on the fly, the magnetic chain (Fig. 3g) generates puller type flow for self-propulsion as observed experimentally and confirmed by simulation (Fig. 4a).[70] For particle-snake hybrid in actuating magnetic fields (Fig. 2a), the flow is pusher type (Fig. 4b).[49] Fig. 4c exhibits the 2-dimensional (2D) flow lines formed by a modular swimmer composed of single IEX and two passive particles with a speed of 1.5 µm/s on the bottom surface in the co-moving frame (top view). The flow is converging but not symmetric in all directions. Especially, at speed ≥ 1.5 µm/s, there is a void region in the back of the swimmer where all tracer particles can not catch up the moving complex. Due to the incompressibility of fluid, the converging flow advects upward and backward.[77] A side view of the asymmetric 3D flow for swimmer shown in Fig. 4c is displayed in the top sketch of Fig. 4d. However, it remains a challenge to detect the 3D flow under swimming, especially the flow between IEX and the colloid, which would be helpful in clarifying the mechanisms of propulsion. According to the conventional classification of swimmers depending on their flow pattern, i.e., pusher, puller or neutral, IEX based modular swimmers belong to none of them. Moreover, the convective flow redistributes the pH gradient in 3D. Therefore, when colloidal particles are lifted up the substrate, the propulsion turns to 3D. The 2D to 3D transition relies on a strong upward convection to overcome the gravity of colloidal particles and circulate them in a convection cell, which has been observed at a large size ratio of IEX to colloid (right photo in Fig. 4e) or increased number of the IEX (left photo in Fig. 4f). However, if passive particles distribute symmetrically, symmetric propulsion leads to eo-pumping of the hybrid cluster (right photo in Fig. 4f).

All the three flow patterns shown in Fig. 4d have been reported for asymmetric dimers (both bound and dynamic) in EHD flow.[44,58] The pattern of flow is determined by the more general types of particle asymmetry (e.g., geometry, surface charge, composition and material) and frequency of the applied electric field inducing the transition of attractive and repulsive EHD flows.[44,58] Also, the swimmers in EHD flow do not belong to pusher, puller or neutral swimmer types.

## 3.2 Types of modular swimming

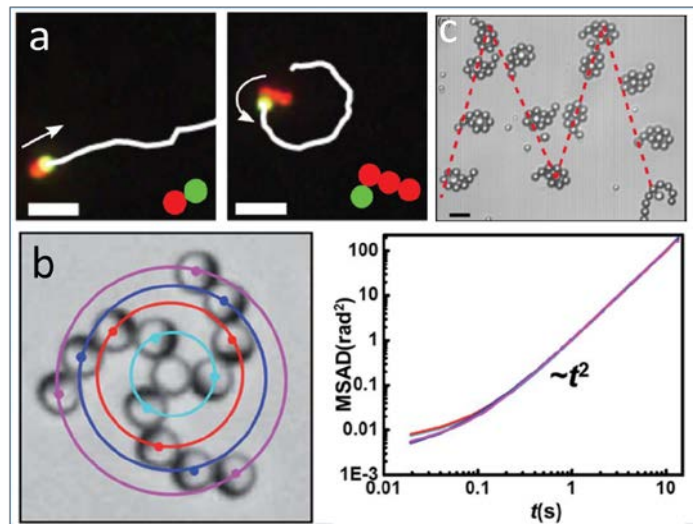

**Fig. 5** Typical types of motion of modular swimmers. (a) Colloidal molecules showing 2D linear and circular motion.[44] (b) Three-arm symmetric pinwheel rotating clockwise, with each particle has the same angular velocity.[57] (c) Transport of a cargo in an "M" shaped trajectory by magnetic lasso.[43] (Reprinted (c) with permission from *Langmuir*, 2017, **33**, 5932. Copyright 2017 American Chemical Society.)

The type of motion performed by modular swimmers depends on the arrangement of the components and is compatible with the motion of each individual constituent. As most modular swimmers are dense compared with the suspension, in experiments, they settle and are confined on the 2 dimensional (2D) plane of bottom substrate. Especially, modular swimmers relying on surface assisted propulsion (e.g., Figs. 1b, 2b, 3d and 3f) have to stay on the surface. For 3D motion, modular swimmers need to rely on other propulsion mechanisms, e.g., bubble thrust.[93,94] As with single unit swimmers, modular microswimmers self-propel linearly on short-time scale when the left-right symmetry relative to the internal propulsion direction is not broken (e.g., Fig. 5a).[95] On long-time scale, Brownian motion randomizes the moving direction of modular swimmers.[39] To keep directed motion, external fields can be applied to bias the disturbance.[50,67] For swimmers with chirality or handedness, the hydrodynamic coupling between translational and rotational motions induces a translated swimmer to rotate, which generates a circular trajectory in 2D (e.g., Fig. 5a).[96] A special case of circular swimmer is the rotating pinwheels, for which the center of rotation is the center of the swimmer. Therefore, in ideal case (without external drift), the pinwheel only has rotational motion with each Janus particle has the same angular velocity and different translational velocities (Fig. 5b).[57] Moreover, modular swimmers can also rearrange or deform under external stimuli. Fig. 5c shows one example of magnetic field induced bending of superparamagnetic chain. We will discuss this point in detail below.

## 4. Steering modular microswimmers

To perform some defined tasks on demand, modular microswimmers need functionality for direction and speed control. For swimmers of single active unit, steering is mainly realized in two ways. First, externally applied uniform fields, e.g., magnetic, electric or ultrasound wave fields were used to bias the moving direction randomized by Brownian motion.[97-99] Second, radically different methods exploiting emergent behavior of individual particles to position dependent stimuli, such as gradient in chemical composition (chemotaxis),[100,101] flow (rheotaxis),[102,103] pH (pH taxis)[21,102] and light (phototaxis)[103] were applied for direction and speed control. In addition, environmental conditions, for example, fuel concentration, reaction kinetics and background salt concentration can also alter the speed of swimmers.[83,106-108]

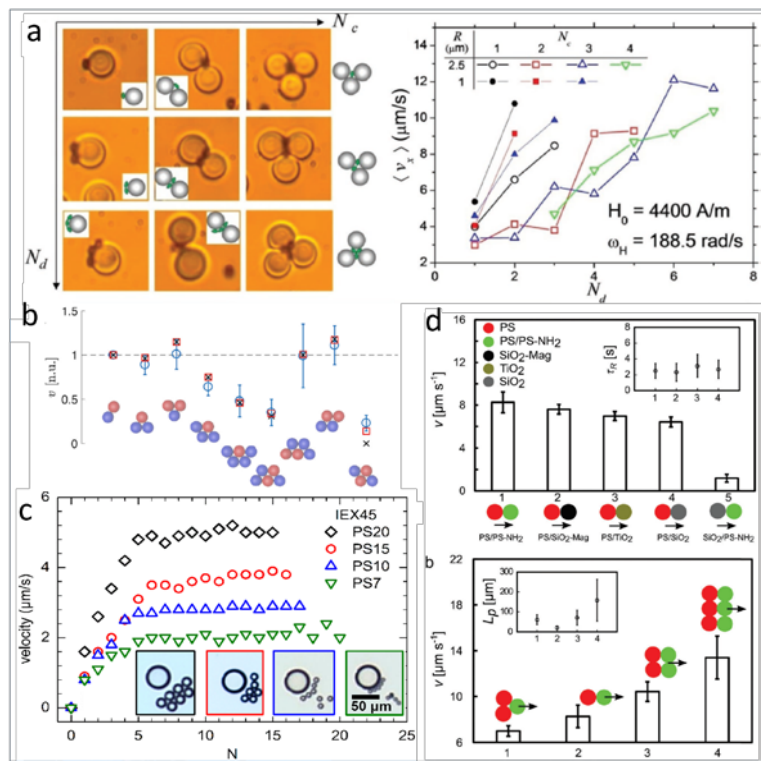

**Fig. 6** Speed control of modular swimmers by composition and arrangement of components. (a) Velocity of swimmers composed of hematite docker and colloidal particles versus number of the attached docker for two differently sized colloid.[66] (b) The speed of swimmers formed by heat adsorbing and non-adsorbing particles of equal size as a function of the composition and arrangement of components.[71] (c) Velocity of modular swimmers formed by single IEX and passive colloids versus number of the colloid of different sizes.[69] (d) The velocity of colloidal molecules with different combinations and number of species propelled by unbalanced EHD flow.[44] (Reprinted (a) from Ref. 66. Copyright © 2017 WILEY.)

The flexible combination of components allow for many functionalities and approaches of steering modular swimmers. For speed control, besides the above mentioned environmental parameters, the more intrinsic parameters for modular swimmers are composition including the number and species (size, material, surface charge etc.) of each component and the arrangement of components. More importantly, dynamic modular swimmers can realize speed control by online compositional change. Martinez-Pedrero et al. reported that the speed of hematite docker-colloid swimmers generally increases with increasing number of the docker and decreasing size of the colloid (Fig. 6a).[66] The number of passive colloid also influences the swimming speed, but the trend is not monotonic. For colloidal molecules of two equally sized spherical particles, the speed decreases with increasing number of heat non-adsorbing particles (Fig. 6b).[71] Furthermore, the arrangement of particles at the same composition is another factor influencing the speed. The above phenomena showing that the speed increases with more drive and decreases with increased friction are intuitive and similar to that observed for cargo transport by swimmers of single active unit.[109,110]

What is counterintuitive is the change of speed of IEX-based modular swimmers with increasing number of passive colloid. The velocity first increases with increasing number of passive particles then saturates when the first row of the IEX is fully filled (Fig. 6c).[69] This trend is robust under varied boundary conditions (background salt concentration, surface charge of the substrate and the colloid, size of the IEX and colloid and the cell height). The increase of speed with increasing number of passive colloid strongly indicates an active role of colloid in self-propulsion. Our phenomenological model explains the mechanism as follows. Each negatively charged particle generates an electrophoretic (ep) flow in the diffusio-electric fields generated by the IEX blowing towards the IEX, which therefore induces the increase of speed.

Since both the fields and flow decay with distance, the weak increase of flow from particles in the second or third rows counterbalances with the increase of friction leading to the saturation of velocity. Consistent with this phenomenological picture, the speed of swimmers also changes with the size ratio between IEX and the colloid. A size ratio of IEX to colloid of about 2 gives the most efficient propulsion, therefore the fastest speed. Theoretical modeling of our modular swimmers is under way.

The speed of asymmetric dimer propelled by unbalanced EHD flow (Figs. 1h and 2i) has been referred to be controlled by the asymmetric dielectric properties, i.e., the zeta potential and Stern layer conductance of the constituents (Fig. 6d).[44,58] Moreover, the speed is tunable by composition.

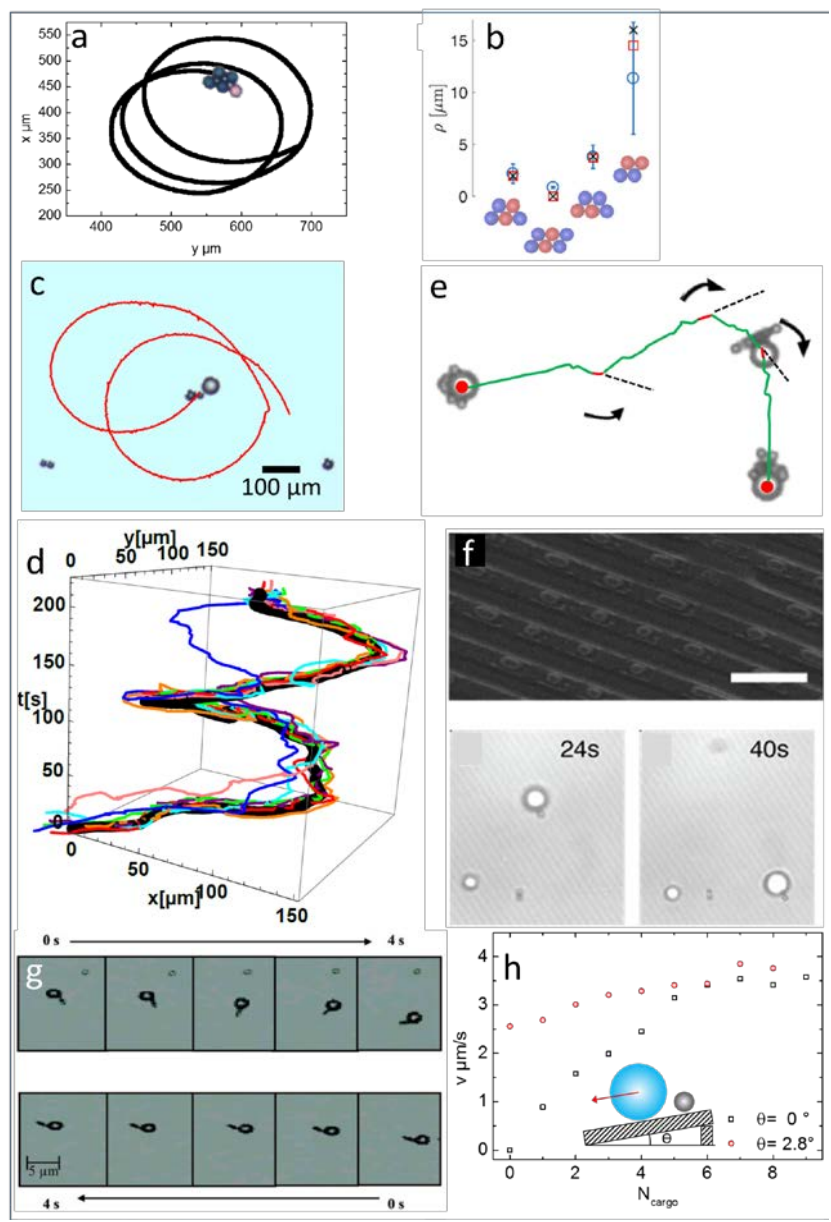

**Fig. 7** Strategies of steering modular swimmers for direction control by changing the arrangement of components (a-e), imprinted microgroove (f),[50] external magnetic (g)[39] and gravitational fields (h). (a) Circular swimmer formed by cationic (red) and anionic IEX (blue) of the same size and asymmetric arrangement overlaid with its trajectory. (b) The radius of circular trajectory of colloidal molecules changes with the number and arrangement of heat adsorbing and non-adsorbing particles.[71] (c) Circular swimmer formed by single IEX and two asymmetric colloidal particles of different sizes overlaid with its trajectory. (d) Trajectories of single IEX (thick black line) and multiple Janus particles (thin colored lines) under modular swimming. (e) Trajectory of modular swimmer composed of single IEX and several superparamagnetic particles with the self-propelling direction controlled by an external magnetic field. (f) Modular swimmer composed by peanut shaped hematite and cargo moves along the microgrooves imprinted on the substrate.[50] (g) Microscopy images showing the trajectory of multi-segmented rod with attached colloidal cargo in 4 s with (lower row) and without (upper row) an external magnetic field.[39] (h) The speed of modular swimmer on a horizontal and a tilted substrates versus number of the colloids. (Reprinted (b) from Ref. 71. Reprinted (f) with permission from *J. Am. Chem. Soc.*, 2013, **135**, 15978. Copyright 2013 American Chemical Society. Reprinted (g) with permission from *Nano Lett.*, 2008, **8**, 1271. Copyright 2008 by American Chemical Society.)

Strategies for direction control can be divided into three categories: structural rearrangement of the swimmer, chemically or physically altered substrates and external field control. Typical examples are shown in Fig. 7. The arrangement of components not only influences the speed but also the moving direction of modular swimmers. We reported that, the linear self-propelling direction of modular swimmers formed by IEX and colloidal particles is along the center of mass of the assembled colloids and the center of IEX.[69] When the center of mass of the assembled colloids and the center of IEX is not in the same line with the internal propulsion direction, a net torque leads to rotational motion. Therefore, when a newly assembled passive particle breaks this symmetry, the swimmer changes from linear to circular swimming. Fig. 7a shows one example of circular swimmers formed by equally sized cationic and anionic IEX when the anionic IEXs organize in an asymmetric way. Similar circular swimmers can also form from equally sized heat adsorbing and non-adsorbing particles as reported by Schmidt et al. (Fig. 3i). Moreover, the authors demonstrated that the radius of the circular trajectory is controllable by the composition and arrangement of components. (Fig. 7b).[71] For IEX-based swimmers, we can actively break the symmetry of the assembled colloids by using asymmetric particles, i.e., particles of different sizes or surface charges. Fig. 7c shows a circular swimmer formed by one IEX (45 µm in diameter) and two differently sized polystyrene particles (diameters of 20 µm and 15 µm). Furthermore, the radius of trajectory is tunable by the size ratio between the colloids and that between the colloid and the IEX or the surface charge ratio between the colloids. Alternatively, a random density fluctuation of the assembled particles would also induce non-linear motion. Fig. 7d shows the trajectory of a swimmer formed by one IEX and a few silica-platinum Janus particles (lower right of Fig. 3f) together with the trajectories of Janus particles at background $H_2O_2$ concentration of 1%. The wobbling motion of Janus particles due to their response to the pH gradient and flow generated by the IEX yields circular swimming. The above approaches for online direction control are feasible; however, relying on self-assembly they are somehow random. A well-controlled approach for structural rearrangement is external magnetic field control when functional magnetic particles are assembled. Fig. 7e shows the trajectory of swimmer formed by single IEX and several superparamagnetic particles manipulated by an external magnetic field. The direction control is realized as follows: when we place a magnetic bar around the swimmer, the superparamagnetic particles form parallel lines along the magnetic field direction. As we rotate the magnet, the lines of particles rotates with the field. When the magnet is removed, the center of mass position of the magnetic particles have already changed relative to the center of the IEX. Consequently, the propelling direction of the swimmer alters.

For modular swimmers with surface assisted propulsion, an alternative way to manipulate the self-propelling direction is chemical or physical alteration of the substrate. This approach has been applied to steer swimmers of single active unit by boundary and patterned substrates[89,111,112] and biological swimmers by microgrooves.[113] For modular swimmers, Palacci et al. showed that the hematite-colloid swimmer moves along the valley of the imprinted microgrooves on the substrate (Fig. 7f).[50] Vilfan et al. showed that the self-propelling direction of rowers (lower picture in Fig. 2g) can also be rectified by circular channels.[55] With the development of 3D printing and microfluidic manufacturing techniques,[114,115] more chemically or physically modified substrates can be designed to guide modular swimmers to specified destinations.

External magnetic/electric field control has become a prominent approach to steering single unit swimmers of asymmetric magnetic properties or anisotropic dielectric properties.[116-119] With implemented functional components, modular swimmers can also be navigated by external fields for direction control. Sundararajan et al. showed that external magnetic field can suppress the rotation of motor-cargo doublets by navigating on the magnetic steering component enabling persistent directed motion (Fig. 7g).[39] For heavy particles, gravitational field is an alternative for direction control. On a tilted surface, modular swimmers formed by IEX and passive cargo switch their self-propelling direction to move downhill. Moreover, the speed of the swimmer gets faster in the velocity increase region owing to the additional gravitational energy converting into motion (Fig. 7h). However, the functions of gravitational field on swimmers of single active unit were reported to be case dependent. Juan et al. reported the downhill rolling of Janus swimmers in viscoelastic solvent.[120] Some bottom heavy particles, e.g., silica-titanium dioxide Janus particles and L shaped swimmers, show negative gravitaxis.[121,122] Therefore, we speculate that uphill swimming is also possible for modular swimmers with suitable designs.

## 5. Challenges and outlook

Owing to the work by several groups, modular swimmers have reached the proof-of-concept stage for realization. They have shown the flexibility in construction, optimization and functionality as well. However, due to the complex balance between different forces, the multiple components and the more available surfaces compared with swimmers of single active unit, theoretical understanding of the mechanism and performance of modular swimmers is a bit challenging, especially those of dynamic structures. There are several theoretical work on modular swimmers of bound structures[123-130] and few on those with dynamic structures.[71,79,80,131] Until now, to the author's knowledge, no microscopic model based on experimental input for modular swimmers of dynamic structures has been proposed. Moving towards the fundamentally interesting and important conditions in terms of potential applications, e.g., crowded environments with more than one swimmers, external gradient (salt, pH, light etc.), viscoelastic fluid or time varying potentials, it is more challenging to predict the performance of modular swimmers. For the case of two interacting modular swimmers, different phenomena have been reported experimentally (Fig. 8). We find that when two colloidal molecules composed of equally sized cationic and anionic IEX are in each other's flow range, they attract each other. After contact, they rearrange and merge into a bigger cluster (Fig. 8a). Depending on the direction of collision, the bigger cluster has different structures and therefore types of motion. However, the interaction between modular swimmer composed of IEX and passive colloids is a bit complex. Within some distance range, two swimmers also attract each other; after contact, they rearrange into a bigger swimmer. However, this swimmer is not stable, it splits into two individual ones. As the two newly formed swimmers are not far from each other, they repeat the approach, merge and split process (Fig. 8b). Zhang et al. reported that two nearby pinwheels synchronize in an external electric field (Fig. 8c).[57] Both careful experimental characterization of the non-equilibrium dynamics and theoretical modeling are needed to reveal the underlying mechanisms. Simulation work showed that catalytic dimers can also move up a concentration gradient of fuel as swimmers of single active unit do.[132] However, no experimental evidence has been reported, especially for dynamic modular swimmers.

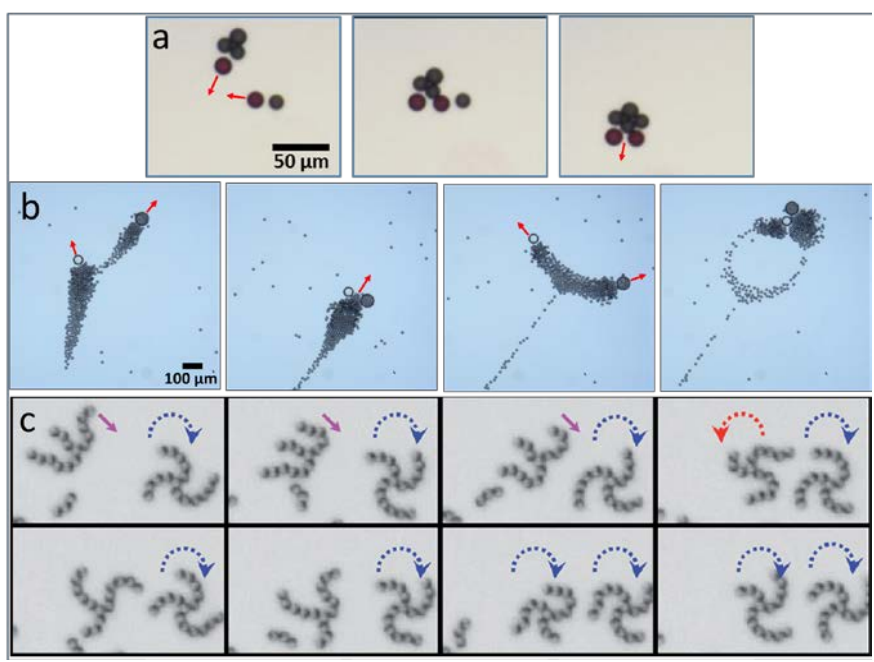

**Fig. 8** Typical examples showing the interaction between modular swimmers. (a) Two self-propelling clusters formed by cationic (red) and anionic IEX (blue) merge into a bigger linearly propelling cluster upon contact. (b) Two IEX-based modular swimmers attract each other, merge, split and remerge. (c) Synchronization of two nearby pinwheel clusters in five seconds.[57] (Reproduced (c) from Ref. 57 with permission from The Royal Society of Chemistry.)

With a large amount of components at hand, more modular swimmers of useful and exciting properties can be constructed making a diverse zoological garden for different specific applications. However, the more important task for current stage is to improve our understanding of modular swimming concerning their mechanism and performance. Therefore, well-defined experiments and theoretical modeling attempts on different levels of approximations are highly needed. Regarding well-defined experiments, the requirement is well-characterized boundary conditions to provide reliable input for theoretical modeling. Taking modular phoretic swimmers as an example, the most important are the gradients (in temperature, pH, solute concentration etc.), the fields driving phoresis (electrostatic, chemical, diffusio-electric etc.), the phoretic mobilities at the involved surfaces and the resulting solvent and solute flows (optimally throughout the complete vessel). Clearly, this presents an experimental challenge. Several clever solutions for specific problems have been reported, which are of great interest for modular swimming. Given the electro-phoretic mobilities of tracers, Farniya et al. managed to infer the underlying diffusio-electric fields in electroosmotic pumping.[133] Using holographic microscopy, several groups were able to accurately monitor swimmer motion and orientation, as well as detect solvent flow fields using buoyancy matched, phoretically inert tracers.[134-136] The distribution of electrolyte concentrations under conditions of inter-diffusion was determined by the fluorescence intensity of dyes added to the diffusing buffer solutions.[137-139] These investigations were carried out in special cells, designed for diffusio-phoretic investigations of passive colloids, and are very interesting to understand the behaviors of passive cargo. In our group, we have developed methods to characterize electro-kinetic mobilities of substrate and cargo and to monitor the development of the underlying pH gradient.[140,141] Regarding theoretical modeling attempts, different theoretical approaches with their expertise ranging from mapping the observations to pair interactions, overexplicit calculation of forces to full treatments of electrostatics, chemical potential and hydrodynamics can be employed.[79,80,123,124,126-130,142,143] With all the efforts, we will reach a new stage with advanced understanding of modular swimming, which in turn may help optimize performance and approach application. We anticipate that modular approach is an alternative step towards designing complex, autonomous and multifunctional micro- and nanomachinery systems.

## Acknowledgement

It is a pleasure to thank Hartmut Löwen, Thomas Speck, Christian Holm, Joost de Graaf, Benno Liebchen, Aidan T Brown and Erdal C. Oğuz for the helpful discussions. We further thank Alexander Reinmüller, Christopher Wittenberg, Denis Botin, Julian Weber, Stanislav Khodorov for their contributions to the project. Internship students Hannah Müller, Songkran Vongsilathi (Thailand) and Zhujun Wu (China) are also acknowledged. We gratefully acknowledge the DFG for financial support (SPP1726, Grants No. PA 459/18-1,2).